\documentclass[12pt]{article}
\UseRawInputEncoding
\usepackage{graphicx}
\begin{document}

\title{Influence of a Galactic Gamma Ray Burst on Ocean Plankton}
\author{}
\maketitle

\centerline{Lien Rodr\'iguez-L\'opez$^1$, Rolando Cardenas$^2$ , Lisdelys Gonz\'alez-Rodr\'iguez$^3$}

\centerline{Mayrene Guimarais$^4$ and J.E. Horvath$^5$}
\bigskip

1- Environmental Science Center (EULA), University of Concepci\'on,

Concepci\'on, Chile

2- Planetary Science Laboratory, Department of  Physics,

UCLV Santa Clara, Cuba

3- Faculty of Engineering, University of Concepci\'on,

Concepci\'on, Chile

4- Engineering Institute, National Autonomous University of Mexico (UNAM), D.F.Mexico

5- Institute of Astronomy, Geophysics and Atmospheric Sciences, University of S\~ao Paulo,

S\~ao Paulo SP, Brazil

\bigskip

\abstract{The hypothesis that one or more biodiversity drops in the Phanerozoic eon, evident in the geological record, might have been caused by the most powerful kind of stellar explosion so far known Gamma Ray Bursts (GRB) has been discussed in several works. These stellar explosions could have left an imprint in the biological evolution on Earth and in other habitable planets. In this work we calculate the short-term lethality that a GRB would produce in the aquatic primary producers on Earth. This effect on life appears because of ultraviolet (UV) retransmission in the atmosphere of a fraction of the gamma energy, resulting in an intense UV flash capable of penetrating tens of meters in the water column in the ocean. We focus on the action of the UV flash on phytoplankton, as they are the main contributors to global aquatic primary productivity. Our results suggest that the UV flash could cause a significant reduction of phytoplankton biomass in the upper mixed layer of the World Ocean.}

\section{Introduction}\label{sec1}

Radiations have the dual role of sterilizing non-resistant species and favoring speciation of the surviving ones due to DNA mutations. Therefore, radiation bursts are plausible hypotheses to explain biodiversity drops and its subsequent increases. One of the natural mechanisms capable of delivering on Earth radiation bursts intense enough are Gamma Ray Bursts (GRB), occurring either in very massive stars or as a consequence of neutron star mergers. For instance, it has been suggested that the GRB could have contributed to the mass extinction Ordovician-Silurian [10,11].

An earlier work [5] compiled and discussed the several effects that a GRB can cause on Earth's atmosphere and biosphere. The best studied is the depletion of the ozone layer, allowing more solar UV to reach the planet's ground during several years. In this work we focus on another important short-term effect: the brief and immediate UV-flash reaching the ground as a result of reprocessing the gamma energy in the atmosphere, and then estimate the immediate lethality that this would cause on phytoplankton. These are the main primary producers and the starting point of the food web in central ocean basins, and are also important in coastal and freshwater ecosystems. Astrophysical calculations based on star formation rate suggest that in the last few billion years each planet in our galaxy would have been affected by a GRB [14, 15, 16]. We thus focus in the short-term lethality that would produce on Earth a typical of the last billion years: a burst arriving from 3000-6000 light years away and delivering 100 $KJ/m^{2}$ of gamma energy at the top of the atmosphere.

\section{Materials and methods}\label{sec2}

\subsection{The interaction of stellar gamma radiation with the atmosphere}

To account for the interaction of the gamma burst with the atmosphere we adopted the results of [9] in this work. The fraction of gamma photons directly reaching the ground is negligible, because of the large Compton cross-section with electrons from the molecules of the atmosphere. The free electrons would then excite other molecules, causing a rich aurora-like spectrum, which reaches the sea level. The ultraviolet fraction of this spectrum (termed the UV-flash) is a major danger for life [5]. The duration of the UV-flash would be the same of the GRBs (around 10 seconds), with a high intensity and even including the very deleterious UV-C band in the wavelength range 260- 280 nm. The interactions of these UV flash photons in water and their efficiency for phytoplankton damage is our concern in this work.

\subsection{Radiative transfer in water and effective doses}

We considered an average ocean albedo of 6.6 \% for zenithal angles not greater than 70 degrees, as reported in Ref.[3]. This was employed to calculate the GRB-UV spectrum just below the ocean surface. We used the classification of optical ocean water types originally presented in [6, 7, 8]. Consequently, we use the attenuation coefficients $K(\lambda)$ of UVR in oceanic water types I, II and III as in [13]. These optical water types can roughly be identified as oligotrophic, mesotrophic and eutrophic, respectively. However, we also included the intermediate types IA and IB. We utilized biological action spectrum for DNA damage $e(\lambda)$ following [3]. Then, the (effective) biological irradiances or dose rates $E*(z)$ at depth $z$ follow from:

\begin{eqnarray}
E^*(z) = \sum e(\lambda)E0(\lambda,0^-)e^{-k(\lambda)z}d\lambda
\end{eqnarray}

The (effective) biological fluences or doses $F*(z)$ are given by:
\begin{eqnarray}
F^*(z) = E^* (z)\Delta t
\end{eqnarray}

where $t$ is the exposure time to UV. We also consider that, just before the UV flash, phytoplankton were homogeneously distributed in the upper mixed layer (UML) of the ocean, due to the mixing action of currents. The depth of UML depends on ocean surface winds and other factors, but after averaging its value for 13 locations [1] we consider it to be 30 meters, quite a typical value.

\subsection{The estimation of induced lethality}
Experiments with phytoplankton stressed by UVR are typically done exposing them to solar radiation during several hours. This is not a scenario close enough to the one we study, given the low intensity of solar UV, as compared to the GRB UV-flash. Therefore, as done by some of us in Ref.[5], we chose the results of Gasc\'on et al. [4]. These authors intensely irradiated a representative set of bacteria with a wavelength ($\lambda= 254 nm$) of the UV-C band. We considered that the more radiation-sensitive phytoplankton would behave as \textit{Escherichia coli}, the intermediate as the aquatic bacterium \textit{Rhodobacter sphaeroides} (wild type and phototrophically grown strain), while the toughest would parallel the soil bacterium \textit{Rhizobium meliloti}. We also analyzed the case in which repair mechanisms would be inhibited: very cold waters or a night-time UV-flash, because at night cell division is synchronized in oceanic phytoplankton [1,2], making them much more radiation sensitive. To account for this last scenario we use the data in [4] for strains in which repair is inhibited due to the lack of recA gene. These data are available for the two extremes of our ``survival band'' (\textit{E. coli} and \textit{R. meliloti}). Strains having the above mentioned gene are denoted \textit{recA+}, while its absence is indicated by \textit{recA-}. We then use the classical model for survival curves of irradiated cells:

\begin{eqnarray}
S = e^{-\alpha F}
\end{eqnarray}

where S is the survival fraction, $\alpha$ is the slope and $F$ is the dose or fluence. However, we introduced some significant refinements. Since the effective biological dose F* calculated from equations (1) and (2) need to be employed, we propose the survival model:
\begin{eqnarray}
S (z) = e^{-(\alpha)F^*}
\end{eqnarray}

where $S(z)$ is the surviving fraction at deep $z$, $\alpha$ is a new (effective biological) slope and $F*(z)$ is the effective biological dose at depth $z$.
The slopes $\alpha$ are a measure of the radiosensitivity of the species. We determine them considering that the reported doses $F$ in [4] follow the simple formula:
\begin{eqnarray}
F = E\Delta t
\end{eqnarray}

Dividing equation (2) by equation (5) we obtain:
\begin{eqnarray}
\frac{F^*}{F}=\frac{E^*}{E}
\end{eqnarray}

Both $F$ and $E$ are given in [4], while $E*$ was determined by Cockell [3] by biologically weighting it, so above equation allows the calculation of $F*$ for each species. We then obtained the biological effective dose for which 10\% of the cells survive $F*_{10}$ using the $F_{10}$ values for each species reported in [4], and finally found the new slope $\alpha$.

\section{Results}\label{sec3}

\subsection{Radiation transfer and effective doses in the ocean}

We used the attenuation coefficients of UVR in the five ocean water types as in [13]. The effective biological doses $F*$ delivered in above mentioned water types are plotted in Fig. 1, as a function of depth $z$. Notice that waters of types I, IA and IB follow a similar behavior.

\begin{figure}[h]
\centerline{\includegraphics[width=10cm]{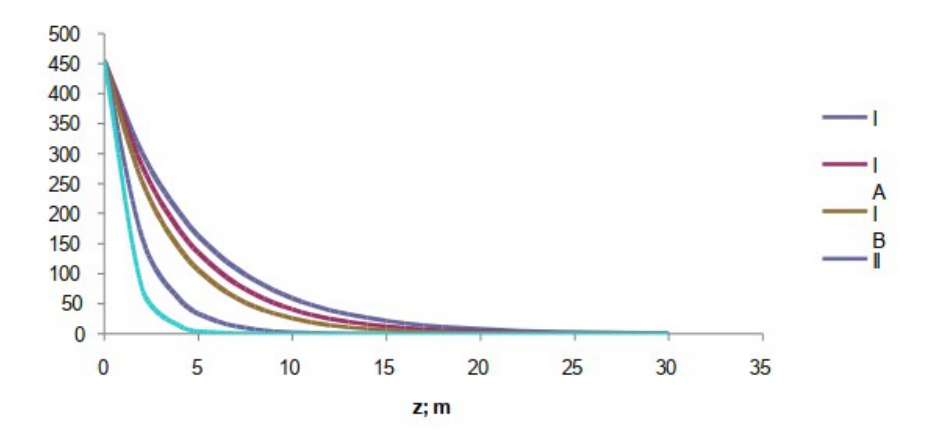}}
\caption{Effective biological doses vs. depth for all ocean optical water types in Jerlov's classification.
\label{fig. 1}}
\end{figure}

\subsection{The estimation of induced lethality}

Here we just show results for the two extremes water types I and III.  Figs. 2 and 3 show the surviving fraction of cells after the GRB UV-flash strikes, for the above mentioned optical ocean water types.

\begin{figure}[h]
\centerline{\includegraphics[width=10cm]{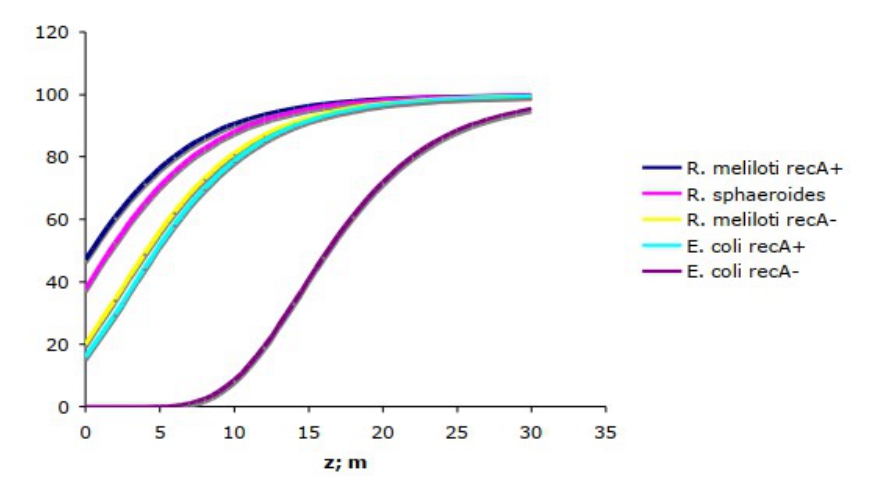}}
\caption{Surviving fraction of cells after the GRB UV-flash strikes, for the case of water type I (clear).
\label{fig. 2}}
\end{figure}

\begin{figure}[h]
\centerline{\includegraphics[width=10cm]{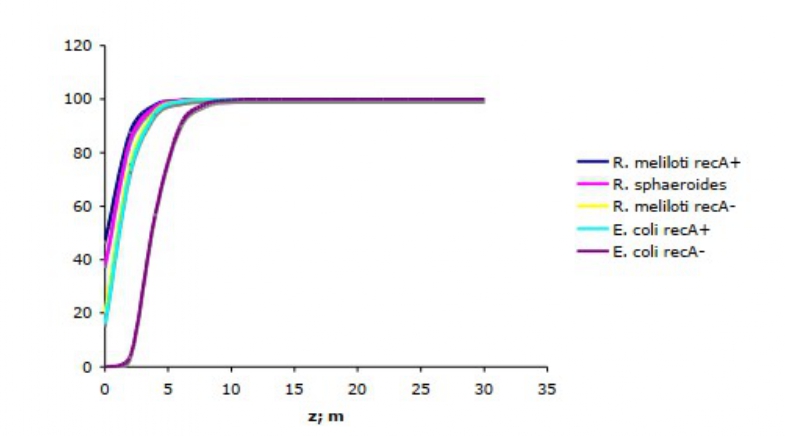}}
\caption{Surviving fraction of cells after the GRB UV-flash strikes, for water type III (turbid).
\label{fig. 3}}
\end{figure}

Table 1 presents the biomass reduction (\% ) in the upper mixed layer of the ocean, for ocean waters I and III.

\begin{tabular}{|c|c|c|c|}
\hline
\textbf{Scenario}                                                                                                  & \textbf{Species}           & \textbf{Type I} & \textbf{Type III} \\
\hline
{Good repair}   & \textit{R. meliloti recA+} & 10,4   & 2,7                       \\
\cline{2-4}     & \textit{R. sphaeroides}    & 12,8   & 3,3                       \\
\cline{2-4}     & \textit{E. coli recA+}     & 20,3   & 5,0                       \\
\hline
Bad repair      & \textit{R. meliloti recA}  & 18,6   & 4,6                       \\
\cline{2-4}     & \textit{E. coli recA-}     & 57,1   & 13,3                      \\
\hline
\end{tabular}

\section{Conclusions}\label{sec4}

Most areas of modern ocean basins are oligotrophic and clear (water types I, IA and IB), thus from Fig. 2 we might expect a significant lethality from a gamma-ray illumination in good repair scenarios, assuming that most species of phytoplankton would behave similarly to the aquatic bacterium \textit{R. sphaeroides}. However, the cells of some species of picoplankton are so small, that it is unlikely that they can host an elaborated DNA repair machinery. An outstanding example is the genus \textit{ Prochlorococcus}, which have been termed the most abundant organisms on Earth, accounting for an estimated 20\% of the oxygen released to the Earth's through the photosynthesis process, and are at the very base of the ocean food assemblage. Given their poor repair capabilities, lethality of species of this genus could be much greater. Figs. 2 and 3 show that phytoplankton living beneath the mixed layer at the moment of the UV-flash would not be affected. Aquatic food webs having a strong dependence on phytoplankton might be very affected and it turns out interesting to evaluate the response of the other primary producers (macroalgae and seagrasses).For the specific scenario of the Ordovician-Silurian extinction, we consider that a GRB it might have been more likely an additive effect  to the glaciation event occurring in the planet. As the World Ocean was predominantly clear, then a rough estimate of phytoplankton biomass reduction (in the upper mixed layer of the ocean) for the whole Earth would be in the range 20-60 \%. This seems to sustain the magnitude of this extinction, the second more severe of the Phanerozoic eon. In a further publication we will present a more detailed modeling of the potential effects of galactic GRB on ocean plankton.

\section*{Acknowledgments}

L.R.-L. says thanks to organizers IWARA 2020 Congress. L.G.-R. acknowledge ANID (Chile) for a PhD Grant. Rolando Cardenas thanks to the Ministry of Higher Education of Cuba for financial support.J.E.H. has been supported by CNPq and FAPESP Agencies (Brazil)

\end{document}